\documentclass[aps,showpacs,twocolumn]{revtex4}

\usepackage{graphics}
\usepackage{epsfig}

\def\ov#1{\overline{#1}}

\def\wt#1{\widetilde{#1}}
\def\vb#1{\mbox{\boldmath$#1$}}
\def\pd#1#2{\frac{\partial #1}{\partial #2}}

\def\wh#1{\widehat{#1}}
\def\bdot{\,\vb{\cdot}\,}
\def\btimes{\,\vb{\times}\,}

\def\bhat{\wh{{\sf b}}}
\def\cal#1{\mathcal{#1}}

\def\exd{{\sf d}}

\newcommand{\bc}{\begin{center}}
\newcommand{\ec}{\end{center}}
\newcommand{\bt}{\begin{tabbing}}
\newcommand{\et}{\end{tabbing}} 
\newcommand{\be}{\begin{eqnarray*}}
\newcommand{\ee}{\end{eqnarray*}}
\newcommand{\bs}{\begin{slide}}
\newcommand{\es}{\end{slide}}

\begin{document}

\title{On the dynamical reduction of the Vlasov equation}

\author{Alain J.~Brizard}
\affiliation{Department of Chemistry and Physics \\ Saint Michael's College, Colchester, VT 05439, USA}

\begin{abstract}
The elimination of a fast time scale from the Vlasov equation by Lie-transform methods is an important step in deriving a reduced Vlasov equation such as the drift-kinetic Vlasov equation or the gyrokinetic Vlasov equation. It is shown here that this dynamical reduction also leads to the introduction of polarization and magnetization effects in the reduced Maxwell equations, which ensure that the reduced Vlasov-Maxwell equations possess an exact energy-momentum conservation law.
\end{abstract}

\begin{flushright}
September 6, 2006
\end{flushright}

\pacs{52.25Dg, 02.40.Yy, 03.50.De}

\maketitle

\section{Introduction}

The asymptotic elimination of fast time scales from the Vlasov equation presents important analytical and computational advantages for its solution in complex plasma geometries \cite{Tang_Chan}. Here, fast time scales in a collisionless plasma are either associated with particle orbital dynamics (e.g., the fast gyromotion of a charged particle about a magnetic field line) or wave-particle dynamics (e.g., the fast oscillatory motion of a charged particle in the presence of a high-frequency electromagnetic wave). In the former case, the elimination of a fast orbital time scale is often associated with the construction of an adiabatic (action) invariant (e.g., the magnetic moment of a charged particle in a strong magnetic field). Next, the construction of a reduced Vlasov equation, from which one or more fast time scales have been removed, can either be obtained through an iterative solution of a perturbatively-expanded Vlasov equation \cite{HTH_67} or by performing one or more near-identity phase-space transformations resulting from applications of Hamiltonian perturbation theory \cite{RGL_82}. The present paper focuses on applications of Lie-transform Hamiltonian perturbation theory.

The most general setting for carrying out Hamiltonian perturbation theory \cite{Brizard_2001} on the Vlasov-Maxwell equations is to use an eight-dimensional extended phase space with coordinates $\cal{Z}^{a} = ({\bf z}; w, t)$, where ${\bf z}$ denotes regular (six-dimensional) phase-space coordinates, $w$ denotes the energy coordinate, and time $t$ denotes its canonically-conjugate coordinate. The extended Hamilton's equations ($\tau$ denotes the Hamiltonian-orbit parameter)
\begin{equation}
\frac{d\cal{Z}^{a}}{d\tau} \;=\; \left\{ \cal{Z}^{a},\; \cal{H}\right\} \;=\; J^{ab}(\cal{Z})\;
\pd{\cal{H}(\cal{Z})}{\cal{Z}^{b}},
\label{eq:extHameq_def}
\end{equation}
are expressed in extended phase space in terms of an extended Hamiltonian $\cal{H}(\cal{Z}) \equiv H({\bf z},t) - w$, where $H({\bf z},t)$ denotes the regular Hamiltonian, and the extended phase-space Lagrangian $\Gamma \equiv \Gamma_{a}\;\exd \cal{Z}^{a}$, which is a differential one-form in extended phase space (summation over repeated indices is, henceforth, implied). Note that the physical motion in extended phase space takes place on the subspace $\cal{H} \equiv 0$:
\begin{equation}
w \;=\; H({\bf z},t).
\label{eq:wH_def}
\end{equation} 
The extended Poisson bracket $\{\;,\;\}$ is obtained from the extended phase-space Lagrangian $\Gamma$, first, by constructing the Lagrange matrix $\vb{\omega}$ (with antisymmetric components $\omega_{ab} \equiv \partial_{a}
\Gamma_{b} - \partial_{b}\Gamma_{a}$) associated with the differential two-form $\omega = \exd\Gamma = 
\frac{1}{2}\,\omega_{ab}\,\exd\cal{Z}^{a}\wedge \exd\cal{Z}^{b}$ and, second, by inverting the Lagrange matrix to obtain the Poisson matrix ${\sf J} \equiv \vb{\omega}^{-1}$ with antisymmetric components $J^{ab} \equiv \{ \cal{Z}^{a},\; 
\cal{Z}^{b}\}$. Hence, we obtain the extended Poisson bracket defined in terms of two arbitrary functions $F$ and $G$ as 
$\{ F,\; G \} \equiv \partial_{a}F\,J^{ab}\,\partial_{b}G$. 

The extended Vlasov equation is expressed in terms of the extended Vlasov distribution $\cal{F}(\cal{Z})$ and the extended Hamilton's equations (\ref{eq:extHameq_def}) as
\begin{equation}
0 \;=\; \frac{d\cal{F}}{d\tau} \;=\; \frac{d\cal{Z}^{a}}{d\tau}\;\pd{\cal{F}(\cal{Z})}{\cal{Z}^{a}} \;\equiv\;  \{ \cal{F},\; \cal{H}\}.
\label{eq:extVlasoveq_def}
\end{equation}
In order to satisfy the physical constraint (\ref{eq:wH_def}), the extended Vlasov distribution is required to be of the form
\begin{equation}
\cal{F}(\cal{Z}) \;\equiv\; c\,\delta[w - H({\bf z}, t)]\; f({\bf z}, t),
\label{eq:F_extended} 
\end{equation}
where $f({\bf z}, t)$ denotes the time-dependent Vlasov distribution on regular phase space. By integrating the extended Vlasov equation (\ref{eq:extVlasoveq_def}) over the energy coordinate $w$ (and using $d\tau = dt$), we obtain the regular Vlasov equation 
\begin{equation}
0 \;=\; \frac{df}{dt} \;\equiv\; \pd{f}{t} \;+\; \{ f,\; H \}. 
\label{eq:reg_Vlasov}
\end{equation}
Note that we use the same symbol $\{\;,\;\}$ for the Poisson bracket on regular phase space in Eq.~(\ref{eq:reg_Vlasov}) since $w$-derivatives appearing in the extended Poisson bracket vanish identically on regular phase space.

Next, the extended Vlasov equation (\ref{eq:extVlasoveq_def}) is coupled with Maxwell's equations for the self-consistent electromagnetic fields
\begin{equation}
\nabla\bdot{\bf E} \;=\; 4\pi\,\rho \;\;\;{\rm and}\;\;\; \nabla\btimes{\bf B} - \frac{1}{c}\pd{{\bf E}}{t} \;=\; \frac{4\pi}{c}\,{\bf J}, 
\label{eq:EB_Maxwell}
\end{equation}
where the charge-current densities
\begin{eqnarray} 
\left( 
\begin{array}{c}
c\rho \\
{\bf J}
\end{array} \right) & = & \sum\;e\;\int d^{4}p\;\cal{F}\;
\left( \begin{array}{c}
c \\
{\bf v}
\end{array} \right) \nonumber \\
 & \equiv & \sum\;e\;\int d^{3}p\;f\;
\left( \begin{array}{c}
c \\
{\bf v}
\end{array} \right)
\label{eq:rhoJ_def}
\end{eqnarray}
are defined in terms of moments of the extended Vlasov distribution $\cal{F}$ (with $d^{4}p = c^{-1}dw\,d^{3}p$) and the electric and magnetic fields ${\bf E} \equiv -\,\nabla\Phi - c^{-1}\partial{\bf A}/\partial t$ and ${\bf B} \equiv \nabla\btimes{\bf A}$ satisfy 
\begin{equation}
\nabla\bdot{\bf B} \;=\; 0 \;\;{\rm and}\;\; \nabla\btimes{\bf E} \;+\; c^{-1}\,\partial{\bf B}/\partial t \;=\; 0.
\label{eq:EB_constraint}
\end{equation}
The purpose of this paper is to show that: the asymptotic elimination of a fast time scale from the Vlasov equation 
(\ref{eq:extVlasoveq_def}) introduces polarization and magnetization effects into the Maxwell equations 
(\ref{eq:EB_Maxwell}); and the reduced Vlasov-Maxwell equations possess exact conservation laws that can be derived from a variational principle by Noether method.

The remainder of this paper is organized as follows. In Sec.~\ref{sec:DR}, we present a brief summary of the Lie-transform perturbation method used in the asymptotic elimination of a fast time scale from the Vlasov equation and its associated Hamiltonian dynamics. In Sec.~\ref{sec:RVM}, we present the reduced Vlasov-Maxwell equations introduced by the dynamical reduction associated with a near-identity phase-space transformation. Through the use of the push-forward representation of particle fluid moments, we present expressions for the charge-current densities involving momentum-space moments of the reduced Vlasov distribution and present explicit expressions for the reduced polarization charge-current densities and the divergenceless reduced magnetization current density. In Sec.~\ref{sec:Var}, we present the variational derivation of the reduced Vlasov-Maxwell equations and derive, through the Noether method, the exact reduced energy-momentum conservation laws (explicit proofs are presented in Appendix \ref{sec:App_A}). Lastly, in Sec.~\ref{sec:Sum}, we summarize the work present here and briefly discuss applications.

\section{\label{sec:DR}Dynamical Reduction by Near-identity Phase-space Transformation}

\subsection{Near-identity Phase-space Transformation}

The process by which a fast time scale is removed from Hamilton's equations (\ref{eq:extHameq_def}) involves a near-identity transformation $\cal{T}_{\epsilon}: \cal{Z} \rightarrow \ov{\cal{Z}}(\cal{Z};\epsilon) \equiv \cal{T}_{\epsilon}\cal{Z}$ on extended particle phase space, where
\begin{equation}
\ov{\cal{Z}}^{a}(\cal{Z},\epsilon) = \cal{Z}^{a} + \epsilon\,G_{1}^{a} + \epsilon^{2} \left( G_{2}^{a} \;+\; \frac{G_{1}^{b}}{2}\,\pd{G_{1}^{a}}{\cal{Z}^{b}} \right) + \cdots,
\label{eq:ovZ_Z}
\end{equation}
and its inverse near-identity transformation $\cal{T}_{\epsilon}^{-1}: \ov{\cal{Z}} \rightarrow \cal{Z}(\ov{\cal{Z}};\epsilon) \equiv \cal{T}_{\epsilon}^{-1}
\ov{\cal{Z}}$, where
\begin{equation}
\cal{Z}^{a}(\ov{\cal{Z}},\epsilon) = \ov{\cal{Z}}^{a} - \epsilon\,G_{1}^{a} - \epsilon^{2} \left( G_{2}^{a} - \frac{G_{1}^{b}}{2}\,\pd{G_{1}^{a}}{\ov{\cal{Z}}^{b}} \right) + \cdots.
\label{eq:Z_ovZ}
\end{equation}
In Eqs.~(\ref{eq:ovZ_Z})-(\ref{eq:Z_ovZ}), the dimensionless ordering parameter $\epsilon \ll 1$ is defined as the ratio of the fast time scale over a slow time scale of interest, and the $n$th-order generating vector field ${\sf G}_{n}$ is chosen to remove the fast time scale at order $\epsilon^{n}$ from the perturbed Hamiltonian dynamics. Examples of asymptotic elimination of fast time scales by Lie-transform perturbation method include guiding-center Hamiltonian theory \cite{RGL_83,Brizard_95}, gyrocenter Hamiltonian theory \cite{BH_06}, and oscillation-center Hamiltonian theory \cite{CK_81}. 

\subsection{Pull-back and Push-forward Operators}

Next, we define the \textit{pull-back} operator on scalar fields \textit{induced} by the near-identity transformation 
(\ref{eq:ovZ_Z}):
\begin{equation}
{\sf T}_{\epsilon}:\; \ov{\cal{F}} \;\rightarrow\; \cal{F} \;\equiv\; {\sf T}_{\epsilon}\ov{\cal{F}},
\label{eq:pullback}
\end{equation}
i.e., the pull-back operator ${\sf T}_{\epsilon}$ transforms a scalar field $\ov{\cal{F}}$ on the phase space with coordinates $\ov{\cal{Z}}$ into a scalar field 
$\cal{F}$ on the phase space with coordinates $\cal{Z}$: 
\[ \cal{F}(\cal{Z}) \;=\; {\sf T}_{\epsilon}\ov{\cal{F}}(\cal{Z}) \;=\; \ov{\cal{F}}(\cal{T}_{\epsilon}\cal{Z}) \;=\; 
\ov{\cal{F}}(\ov{\cal{Z}}). \]
Using the inverse transformation (\ref{eq:Z_ovZ}), we also define the {\it push-forward} operator: 
\begin{equation}
{\sf T}_{\epsilon}^{-1}:\; \cal{F} \;\rightarrow\; \ov{\cal{F}} \;\equiv\; {\sf T}_{\epsilon}^{-1}\cal{F},
\label{eq:pushforward}
\end{equation}
i.e., the push-forward operator ${\sf T}_{\epsilon}^{-1}$ transforms a scalar field $\cal{F}$ on the phase space with coordinates $\cal{Z}$ into a scalar field 
$\ov{\cal{F}}$ on the phase space with coordinates $\ov{\cal{Z}}$: 
\[ \ov{\cal{F}}(\ov{\cal{Z}}) \;=\; {\sf T}_{\epsilon}^{-1}\cal{F}(\ov{\cal{Z}}) \;=\; \cal{F}({\cal T}_{\epsilon}^{-1}
\ov{\cal{Z}}) \;=\; \cal{F}(\cal{Z}). \]
Note that both induced transformations (\ref{eq:pullback}) and (\ref{eq:pushforward}) satisfy the scalar-invariance property $\cal{F}(\cal{Z}) = \ov{\cal{F}}(\ov{\cal{Z}})$. 

In Lie-transform perturbation theory \cite{RGL_82}, the pull-back and push-forward operators (\ref{eq:pullback}) and (\ref{eq:pushforward}) are expressed as Lie 
{\it transforms}: ${\sf T}_{\epsilon}^{\pm 1} \equiv \exp(\pm\sum_{n = 1}\epsilon^{n}\,\pounds_{n})$ defined in terms of the Lie {\it derivative} $\pounds_{n}$ generated by the $n$th-order vector field ${\sf G}_{n}$, which appear in the 
$n$th-order terms found in the near-identities (\ref{eq:ovZ_Z}) and (\ref{eq:Z_ovZ}). A Lie derivative is a special differential operator that preserves the tensorial nature of the object it operates on \cite{AM_78}. For example, the Lie derivative $\pounds_{n}$ of the scalar field $\cal{H}$ is defined as the scalar field $\pounds_{n}\cal{H} \equiv G_{n}^{a}\,\partial_{a}\cal{H}$, while the Lie derivative $\pounds_{n}$ of a one-form $\Gamma \equiv \Gamma_{a}\,\exd\cal{Z}^{a}$ is defined as the one-form 
\[ \pounds_{n}\Gamma \;\equiv\; G_{n}^{a}\,\omega_{ab}\,\exd\cal{Z}^{b} \;+\; \exd( G_{n}^{a}\;\Gamma_{a}), \]
where $\omega_{ab} \equiv \partial_{a}\Gamma_{b} - \partial_{b}\Gamma_{a}$ are the components of the two-form 
$\vb{\omega} \equiv \exd\Gamma$.

The pull-back and push-forward operators (\ref{eq:pullback})-(\ref{eq:pushforward}) can now be used to transform an arbitrary operator $\cal{C}: F(\cal{Z}) \rightarrow \cal{C}[F](\cal{Z})$ acting on the extended Vlasov distribution function $\cal{F}$. First, since $\cal{C}[\cal{F}](\cal{Z})$ is a scalar field, it transforms to ${\sf T}_{\epsilon}^{-1}\{\cal{C}[\cal{F}]\}(\ov{\cal{Z}})$ with the help of the push-forward operator (\ref{eq:pushforward}). Next, we replace the extended Vlasov distribution function $\cal{F}$ with its pull-back representation $\cal{F} = {\sf T}_{\epsilon}\ov{\cal{F}}$ and, thus, we define the transformed operator
\begin{equation}
\cal{C}_{\epsilon}[\ov{\cal{F}}] \;\equiv\; {\sf T}_{\epsilon}^{-1}(\cal{C}[{\sf T}_{\epsilon}\ov{\cal{F}}]). 
\label{eq:redC}
\end{equation}
By applying the induced transformation (\ref{eq:redC}) on the extended Vlasov operator $d/d\tau$ defined in 
Eq.~(\ref{eq:extVlasoveq_def}), we obtain
\begin{equation}
\frac{d_{\epsilon}\ov{F}}{d\tau} \;\equiv\; {\sf T}_{\epsilon}^{-1}\left( 
\frac{d}{d\tau}\;{\sf T}_{\epsilon}\ov{F} \right) \;\equiv\; \{ \ov{F},\; \ov{\cal{H}} \}_{\epsilon},
\label{eq:extVlasov_new}
\end{equation} 
where the total derivative $d_{\epsilon}/d\tau$ along the transformed particle orbit is defined in terms of the transformed Hamiltonian
\begin{equation}
\ov{\cal{H}} \;\equiv\; {\sf T}_{\epsilon}^{-1}\cal{H},
\label{eq:eham_lt}
\end{equation} 
and the transformed Poisson bracket 
\begin{equation} 
\{ \ov{F},\; \ov{G} \}_{\epsilon} \;\equiv\; {\sf T}_{\epsilon}^{-1}( \{ {\sf T}_{\epsilon}\ov{F},\; 
{\sf T}_{\epsilon}\ov{G} \}).
\label{eq:redPB}
\end{equation}
The Poisson-bracket transformation $\{\;,\;\} \rightarrow \{\;,\;\}_{\epsilon}$ can also be performed through the transformation of the extended phase-space Lagrangian, $\ov{\Gamma}_{\epsilon} \equiv {\sf T}_{\epsilon}^{-1}\Gamma + \exd\cal{S}$, is expressed as \cite{RGL_82}
\begin{eqnarray*} 
 &  &\ov{\Gamma}_{\epsilon} \;=\; \Gamma_{0} \;+\; \epsilon \left( \Gamma_{1} \;+\; \exd\cal{S}_{1} \;-\; \pounds_{1}\Gamma_{0} \right) \\
 &  &+\; \epsilon^{2} \left( \Gamma_{2} + \exd\cal{S}_{2} - \pounds_{2}\Gamma_{0} - \pounds_{1}\Gamma_{1} + \frac{1}{2}\, \pounds_{1}^{2}\Gamma_{0} \right) + \cdots,
\end{eqnarray*}
where $\cal{S} \equiv \epsilon\,\cal{S}_{1} + \epsilon^{2}\,\cal{S}_{2} + \cdots$ denotes a (canonical) scalar field used to simplify the transformed phase-space Lagrangian $\ov{\Gamma}_{\epsilon}$ at each order $\epsilon^{n}$ in the perturbation analysis. Note that the choice of $\cal{S}$ has no impact on the new Poisson-bracket structure 
\[ \ov{\omega}_{\epsilon} \;=\; \exd\ov{\Gamma}_{\epsilon} \;=\; \exd\left({\sf T}_{\epsilon}^{-1}\Gamma\right) \;=\; 
{\sf T}_{\epsilon}^{-1}\exd\Gamma \;\equiv\; {\sf T}_{\epsilon}^{-1}\omega, \]
since $\exd^{2}\cal{S} = 0$ (i.e., $\partial^{2}_{ab}\cal{S} - \partial^{2}_{ba}\cal{S} = 0$) and ${\sf T}_{\epsilon}^{-1}$ commutes with $\exd$. By inverting the reduced Lagrange matrix $\ov{\omega}_{\epsilon} = {\sf T}_{\epsilon}^{-1}\omega \rightarrow \ov{{\sf J}}_{\epsilon} \equiv \ov{\vb{\omega}}_{\epsilon}^{-1}$, we, thus, obtain the reduced Poisson matrix $\ov{{\sf J}}_{\epsilon}$, with antisymmetric components $\ov{J}_{\epsilon}^{ab} \equiv \{ \ov{\cal{Z}}^{a},\; 
\ov{\cal{Z}}^{b}\}_{\epsilon}$, and define the reduced Poisson bracket $\{ \ov{F},\; \ov{G} \}_{\epsilon} \equiv 
\partial_{a}\ov{F}\;\ov{J}_{\epsilon}^{ab}\;\partial_{b}\ov{G}$. Lastly, we note that the extended-Hamiltonian transformation (\ref{eq:eham_lt}) may be re-expressed in terms of the regular Hamiltonians $H$ and $\ov{H} \equiv 
{\sf T}_{\epsilon}^{-1}H - \partial\cal{S}/\partial t$ as \cite{RGL_82}
\begin{eqnarray*}
 &  &\ov{H} \;=\; H_{0} \;+\; \epsilon \left( H_{1} \;-\; \pounds_{1}H_{0} \;-\; \pd{\cal{S}_{1}}{t} \right) \\
 &  &+\; \epsilon^{2} \left( H_{2} - \pounds_{2}H_{0} - \pounds_{1}H_{1} + \frac{1}{2}\,\pounds_{1}^{2}H_{0} - 
\pd{\cal{S}_{2}}{t} \right) + \cdots.
\end{eqnarray*}
The new extended phase-space coordinates are chosen (i.e., the generating vector field ${\sf G}_{n}$ and the scalar field $\cal{S}_{n}$ are specified at each order $n = 1, 2, ...$ in the perturbation analysis) so that $d_{\epsilon}\ov{\cal{Z}}^{a}/d\tau = \{\ov{\cal{Z}}^{a},\; \ov{\cal{H}}\}_{\epsilon}$ are independent of the fast time scale.

\section{\label{sec:RVM}Reduced Vlasov-Maxwell Equations}

\subsection{Reduced Vlasov Equation}

The push-forward transformation of the extended Vlasov distribution (\ref{eq:F_extended}) yields the reduced 
extended Vlasov distribution 
\begin{equation}
\ov{\cal{F}}(\ov{\cal{Z}}) \;\equiv\; c\,\delta[\ov{w} - \ov{H}(\ov{{\bf z}}, t)]\; \ov{f}(\ov{{\bf z}}, t),
\label{eq:ovF_extended} 
\end{equation}
where the reduced extended Hamiltonian $\ov{\cal{H}} \equiv \ov{H}(\ov{{\bf z}}, t) - \ov{w}$ is defined in Eq.~(\ref{eq:eham_lt}). The extended reduced Vlasov equation 
\begin{equation}
\frac{d_{\epsilon}\ov{\cal{F}}}{d\tau} \;\equiv\; \{ \ov{\cal{F}},\; \ov{\cal{H}}\}_{\epsilon} \;=\; 0
\label{eq:Vlasov_epsilon}
\end{equation}
can be converted into the regular reduced Vlasov equation by integrating it over the reduced energy coordinate $\ov{w}$, which yields the reduced Vlasov equation
\begin{equation}
0 \;=\; \frac{d_{\epsilon}\ov{f}}{dt} \;\equiv\; \pd{\ov{f}}{t} \;+\; \{ \ov{f},\; \ov{H}\}_{\epsilon},
\label{eq:redVlasov_def} 
\end{equation}
where $\ov{f}(\ov{{\bf z}},t)$ denotes the time-dependent reduced Vlasov distribution on the new reduced phase space. Hence, we see that the pull-back and push-forward operators play a fundamental role in the transformation of the Vlasov equation to the reduced Vlasov equation. 

\subsection{Reduced Maxwell Equations}

We now investigate how the pull-back and push-forward operators (\ref{eq:pullback}) and (\ref{eq:pushforward}) are used in the transformation of Maxwell's equations (\ref{eq:EB_Maxwell}). The charge-current densities (\ref{eq:rhoJ_def}) can be expressed in terms of the general expression (where time dependence is omitted for clarity)
\begin{equation} 
J^{\mu}({\bf r}) \;=\; \sum\;e\;\int d^{3}x\,\int d^{4}p\;v^{\mu}\,\delta^{3}({\bf x} - {\bf r})\;\cal{F},
\label{eq:chi_mu}
\end{equation}
where the delta function $\delta^{3}({\bf x} - {\bf r})$ means that only particles whose positions ${\bf x}$ coincide with the field position ${\bf r}$ contribute to the moment $J^{\mu}({\bf r})$. By applying the extended (time-dependent) phase-space transformation $\cal{T}_{\epsilon}:\,\cal{Z} \rightarrow \ov{\cal{Z}}$ (where time $t$ itself is unaffected) on the right side of Eq.~(\ref{eq:chi_mu}), we obtain the push-forward representation for $J^{\mu}$:
\begin{eqnarray} 
J^{\mu}({\bf r}) & = & \sum\;e\;\int d^{3}\ov{x}\,d^{4}\ov{p}\;\left( {\sf T}_{\epsilon}^{-1}v^{\mu}\right)\,
\delta^{3}(\ov{{\bf x}} + \vb{\rho}_{\epsilon} - {\bf r})\;\ov{\cal{F}} \nonumber \\
 & = & \sum\;e\;\int d^{4}\ov{p}\; \left[ \frac{}{} \left( {\sf T}_{\epsilon}^{-1}v^{\mu}\right)\,\ov{\cal{F}} \right. \nonumber \\
 &  &\left.\hspace*{0.5in}-\; \nabla\bdot\left( \vb{\rho}_{\epsilon}\,{\sf T}_{\epsilon}^{-1}v^{\mu}\;\ov{\cal{F}} \right) + \cdots \frac{}{} \right],
\label{eq:chimu_epsilon}
\end{eqnarray}
where ${\sf T}_{\epsilon}^{-1}v^{\mu} = (c, {\sf T}_{\epsilon}^{-1}{\bf v})$ denotes the push-forward of the particle four-velocity $v^{\mu}$ and the displacement $\vb{\rho}_{\epsilon} \equiv {\sf T}_{\epsilon}^{-1}{\bf x} - \ov{{\bf x}}$ between the push-forward ${\sf T}_{\epsilon}^{-1}{\bf x}$ of the particle position ${\bf x}$ and the (new) reduced position $\ov{{\bf x}}$ is expressed as
\begin{equation}
\vb{\rho}_{\epsilon} \;=\; -\,\epsilon\;G_{1}^{{\bf x}} \,-\, \epsilon^{2} \left( G_{2}^{{\bf x}} - \frac{1}{2}\,
{\sf G}_{1}\cdot\exd G_{1}^{{\bf x}} \right) + \cdots
\label{eq:rhoepsilon_def}
\end{equation}
in terms of the generating vector fields $({\sf G}_{1},{\sf G}_{2}, \cdots)$ associated with the near-identity transformation. 

The push-forward representation for the charge-current densities, therefore, naturally introduces polarization and magnetization effects into the Maxwell equations. Hence, the microscopic Maxwell's equations (\ref{eq:EB_Maxwell}) are transformed into the macroscopic (reduced) Maxwell equations
\begin{eqnarray}
\nabla\bdot{\bf D} & = & 4\pi\,\ov{\rho}, \label{eq:D_Poisson} \\
\nabla\btimes{\bf H} \;-\; \frac{1}{c}\,\pd{{\bf D}}{t} & = & \frac{4\pi}{c}\,\ov{{\bf J}},
\label{eq:H_Ampere}
\end{eqnarray} 
where the reduced charge-current densities $\ov{J}^{\mu} \equiv (c\ov{\rho}, \ov{{\bf J}})$ are defined as moments of the reduced Vlasov distribution $\ov{\cal{F}}$: 
\begin{equation} 
\left( c\ov{\rho}, \ov{{\bf J}} \right) \;=\; \sum\;e\;\int d^{4}\ov{p}\;\ov{\cal{F}}\;
\left( c,\; \frac{d_{\epsilon}\ov{{\bf x}}}{dt} \right),
\label{eq:RedrhoJ_def}
\end{equation}
where $d_{\epsilon}\ov{{\bf x}}/dt$ denotes the reduced (e.g., guiding-center) velocity. The microscopic electric and magnetic fields ${\bf E}$ and ${\bf B}$ are, thus, replaced by the macroscopic fields 
\begin{equation}
\left. \begin{array}{rcl}
{\bf D} & = & {\bf E} \;+\; 4\pi\,{\bf P}_{\epsilon} \\
{\bf H} & = & {\bf B} \;-\; 4\pi\,{\bf M}_{\epsilon}
\end{array} \right\},
\label{eq:DH_def}
\end{equation}
where ${\bf P}_{\epsilon}$ and ${\bf M}_{\epsilon}$ denote the polarization and magnetization vectors associated with the dynamical reduction introduced by the phase-space transformation (\ref{eq:ovZ_Z}). 

\subsection{Push-forward Representation of Charge-Current Densities}

We now derive explicit expressions for the reduced polarization ${\bf P}_{\epsilon}$ and the reduced magnetization ${\bf M}_{\epsilon}$ by using the push-forward representation method. First, we derive the push-forward representation (\ref{eq:chimu_epsilon}) for the charge density $(J^{0}= c\rho)$:
\begin{eqnarray}
\rho & = & \sum\;e\; \int d^{4}\ov{p}\;\ov{\cal{F}} \;-\; \nabla\bdot\left( \sum\;e\; \int d^{4}\ov{p}\;
\vb{\rho}_{\epsilon}\;\ov{\cal{F}} \;+\; \cdots \right) \nonumber \\
 & \equiv & \ov{\rho} \;-\; \nabla\bdot{\bf P}_{\epsilon},
\label{eq:rho_def}
\end{eqnarray} 
where $\ov{\rho} \equiv \sum\,e\,\int d^{4}\ov{p}\,\ov{\cal{F}}$ denotes the reduced charge density that appears in Eq.~(\ref{eq:D_Poisson}) and the polarization vector is defined as
\begin{equation}
{\bf P}_{\epsilon} \;\equiv\; \sum\;e \int d^{4}\ov{p} \left[\; \vb{\rho}_{\epsilon}\;\ov{\cal{F}} \;-\; \frac{1}{2}\,
\nabla\bdot \left( \frac{}{} \vb{\rho}_{\epsilon}\,\vb{\rho}_{\epsilon}\;\ov{\cal{F}} \right) + \cdots \right],
\label{eq:Pepsilon_def}
\end{equation}
where the quadrupole contribution $\frac{1}{2}\,\nabla\bdot(\vb{\rho}_{\epsilon}\vb{\rho}_{\epsilon}\,\ov{\cal{F}})$ (which will be useful in what follows) is retained in Eq.~(\ref{eq:Pepsilon_def}), while the reduced electric-dipole 
moment (for each particle species)
\begin{equation}
\vb{\pi}_{\epsilon} \;\equiv\; e\;\ov{\vb{\rho}}_{\epsilon} 
\label{eq:pi_def}
\end{equation}
is associated with the fast-time-averaged charge separation induced by the near-identity phase-space transformation. 

Secondly, we derive the push-forward expression for the current density ${\bf J}$, where the push-forward of the particle velocity ${\bf v} = d{\bf x}/dt$ (using the Lagrangian representation)
\begin{eqnarray} 
{\sf T}_{\epsilon}^{-1}{\bf v} & = & {\sf T}_{\epsilon}^{-1}\frac{d{\bf x}}{dt} \;=\; \left[ {\sf T}_{\epsilon}^{-1}
\frac{d}{dt}{\sf T}_{\epsilon}\right]\left({\sf T}_{\epsilon}^{-1}{\bf x}\right) \nonumber \\
 & \equiv & \frac{d_{\epsilon}\ov{{\bf x}}}{dt} \;+\; \frac{d_{\epsilon}\vb{\rho}_{\epsilon}}{dt}
\label{eq:Tminus_v}
\end{eqnarray}
is expressed in terms of the reduced velocity $d_{\epsilon}\ov{{\bf x}}/dt$, which is independent of the fast time scale, and the particle {\it polarization} velocity $d_{\epsilon}\vb{\rho}_{\epsilon}/dt$, which has both fast and slow time dependence. Note that the fast-time-average particle {\it polarization} velocity 
$d_{\epsilon}\ov{\vb{\rho}}_{\epsilon}/dt$, which is nonvanishing under certain conditions, represents additional reduced dynamical effects (e.g., the standard polarization drift in guiding-center theory \cite{Sosenko,SBD}) not included in $d_{\epsilon}\ov{{\bf x}}/dt$. Hence, the push-forward expression 
(\ref{eq:chimu_epsilon}) for the current density ${\bf J}$ is
\begin{eqnarray}
 &  &{\bf J} \;=\; \sum\;e\; \int d^{4}\ov{p}\;\left( \frac{d_{\epsilon}\ov{{\bf x}}}{dt} \;+\; 
\frac{d_{\epsilon}\vb{\rho}_{\epsilon}}{dt} \right) \ov{\cal{F}} \nonumber \\
 &  &-\; \nabla\bdot\left[ \sum\,e \int d^{4}\ov{p}\;\vb{\rho}_{\epsilon}\,\left( \frac{d_{\epsilon}\ov{{\bf x}}}{dt} + \frac{d_{\epsilon}\vb{\rho}_{\epsilon}}{dt} \right) \ov{\cal{F}} \right] + \cdots,
\label{eq:J_def}
\end{eqnarray}
We may now replace the polarization velocity $d_{\epsilon}\vb{\rho}_{\epsilon}/dt$ in Eq.~(\ref{eq:J_def}) by using the following identity based on the reduced polarization vector (\ref{eq:Pepsilon_def}):
\begin{eqnarray}
\pd{{\bf P}_{\epsilon}}{t} & = & \sum\;e\; \int d^{4}\ov{p}\;\left(\frac{d_{\epsilon}\vb{\rho}_{\epsilon}}{dt}\right)\;
\ov{\cal{F}} \nonumber \\
 &  &-\; \nabla\bdot \left\{ \sum\;e\; \int d^{4}\ov{p}\;\ov{\cal{F}} \left[\; \left(\frac{d_{\epsilon}\ov{{\bf x}}}{dt}\right)\;\vb{\rho}_{\epsilon} \right. \right. \nonumber \\
 &  &\left. \left. \hspace*{0.5in}+\; \frac{1}{2}\;\frac{d_{\epsilon}}{dt}\left( \frac{}{} \vb{\rho}_{\epsilon}\,\vb{\rho}_{\epsilon} \right) \;+\; \cdots \;\right] \right\},
\label{eq:Pepsilon_time}
\end{eqnarray}
where the reduced Vlasov equation (\ref{eq:extVlasov_new}) was used and integration by parts was performed. Using the vector identity $\nabla\bdot({\bf B}{\bf A} - {\bf A}{\bf B}) \equiv \nabla\btimes({\bf A}\btimes{\bf B})$, the push-forward representation for the current density is, therefore, expressed as
\begin{equation}
{\bf J} \;\equiv\; \ov{{\bf J}} \;+\; \pd{{\bf P}_{\epsilon}}{t} \;+\; c\;\nabla\btimes{\bf M}_{\epsilon},
\label{eq:Jepsilon_def}
\end{equation}
where $\ov{{\bf J}} \equiv \sum\,e\,\int d^{4}\ov{p}\,(d_{\epsilon}\ov{{\bf x}}/dt)\,\ov{\cal{F}}$ denotes the reduced current density appearing in Eq.~(\ref{eq:H_Ampere}), ${\bf J}_{{\rm pol}} \equiv \partial{\bf P}_{\epsilon}/\partial t$ denotes the reduced polarization current, and the divergenceless reduced magnetization current ${\bf J}_{{\rm mag}} \equiv c\,\nabla\btimes{\bf M}_{\epsilon}$ is expressed in terms of the reduced magnetization vector
\begin{eqnarray}
{\bf M}_{\epsilon} & = & \sum\;\frac{e}{c}\;\int d^{4}\ov{p}\; \vb{\rho}_{\epsilon}\btimes\left( \frac{1}{2}\;\frac{d_{\epsilon}\vb{\rho}_{\epsilon}}{dt} \;+\;
\frac{d_{\epsilon}\ov{{\bf x}}}{dt} \right) \ov{\cal{F}} \nonumber \\
 & \equiv & \sum\;\int d^{4}\ov{p}\; \left( \vb{\mu}_{\epsilon} \;+\; \frac{\vb{\pi}_{\epsilon}}{c}\btimes
\frac{d_{\epsilon}\ov{{\bf x}}}{dt} \right) \ov{\cal{F}},
\label{eq:Mepsilon_def}
\end{eqnarray}
which represents the sum (for each particle species) of the intrinsic (fast-time-averaged) magnetic-dipole moment
\begin{equation}
\vb{\mu}_{\epsilon} \;\equiv\; \frac{e}{2c}\;\ov{\left(\vb{\rho}_{\epsilon}\btimes \frac{d_{\epsilon}
\vb{\rho}_{\epsilon}}{dt}\right)},
\label{eq:mu_def}
\end{equation}
and a moving electric-dipole contribution $(\vb{\pi}_{\epsilon}\btimes d_{\epsilon}\ov{{\bf x}}/dt)$, as suggested by classical electromagnetic theory \cite{Jackson}.

\section{\label{sec:Var}Variational Formulation of Reduced Vlasov-Maxwell Equations}

\subsection{Reduced Variational Principle}

We now show that the reduced Vlasov-Maxwell equations (\ref{eq:Vlasov_epsilon}) and (\ref{eq:D_Poisson})-(\ref{eq:H_Ampere}) can be derived from a variational principle $\int d^{4}x\;\delta\ov{\cal{L}} = 0$. Variational principles for reduced Vlasov-Maxwell equations have been presented previously by Pfirsch \cite{Pfirsch} and Pfirsch and Morrison \cite{PM_85} using the Hamilton-Jacobi formulation and by Kaufman {\it et al.} \cite{Kaufman_84,KH_84,PLS,Boghosian,Ye_K} using the Low-Lagrangian formalism. Here, we present a variational principle for reduced Vlasov-Maxwell equations based on the reduced Lagrangian density \cite{Brizard_2000}
\begin{equation}
\ov{\cal{L}} \;\equiv\; \frac{{\sf F}:{\sf F}}{16\pi} \;-\; \sum\;\int d^{4}\ov{p}\;\ov{\cal{F}}\;\ov{\cal{H}},
\label{eq:redLag_def}
\end{equation}
where $\ov{\cal{F}}$ and $\ov{\cal{H}}$ denote the reduced extended Vlasov distribution and the reduced extended Hamiltonian, respectively. In this Section, we use the convenient space-time metric $g^{\mu\nu} = {\rm diag}\,(-1, 1, 1, 1)$, so that the electromagnetic field tensor $F_{\mu\nu} \equiv \partial_{\mu}A_{\nu} - \partial_{\nu}A_{\mu}$ is defined in terms of the four-potential $A_{\mu} = (-\,\Phi, {\bf A})$ and, thus, $F_{i0} = E_{i}$ and $F_{ij} = \epsilon_{ijk}\,B^{k}$. In order to simplify our presentation, we use canonical four-momentum coordinates $\ov{p}_{\mu} = (-\ov{w}/c, \ov{{\bf p}})$, so that the reduced extended Hamiltonian $\ov{\cal{H}}$ is required to be invariant under the gauge transformation 
\begin{equation}
A_{\mu} \;\rightarrow\; A_{\mu} \;+\; \partial_{\mu}\chi \;\;\;{\rm and}\;\;\; \ov{p}_{\mu} \;\rightarrow\; \ov{p}_{\mu} \;+\; (e/c)\,\partial_{\mu}\chi, 
\label{eq:gauge}
\end{equation}
where $\chi$ denotes the electromagnetic gauge field.

Note that, as a result of the dynamical reduction of the Vlasov equation, the reduced Hamiltonian $\ov{\cal{H}}$ is not only a function of the four-potential $A_{\mu}$ but also of the field tensor $F_{\mu\nu}$. From these dependences, we express the reduced four-current density (\ref{eq:RedrhoJ_def}) as
\begin{equation}
\ov{J}^{\mu} \;=\; (c\ov{\rho}, \ov{{\bf J}}) \;\equiv\; -\;c\;\sum\;\int d^{4}\ov{p}\;\ov{\cal{F}}\;\pd{\ov{\cal{H}}}{A_{\mu}},
\label{eq:redJ_var}
\end{equation}
and we introduce the reduced antisymmetric polarization-magnetization tensor \cite{PLS}
\begin{equation}
K^{\mu\nu} \;\equiv\; -\;\sum\;\int d^{4}\ov{p}\;\ov{\cal{F}}\;\pd{\ov{\cal{H}}}{F_{\mu\nu}},
\label{eq:redH_var}
\end{equation}
where the reduced polarization and magnetization vectors $K^{i0} = P_{\epsilon}^{i}$ and $K^{ij} = \epsilon^{ijk}\,
M_{\epsilon\,k}$ are defined as  
\begin{equation}
\left( {\bf P}_{\epsilon},\; {\bf M}_{\epsilon} \right) \;\equiv\; -\;\sum\;\int d^{4}\ov{p}\;\ov{\cal{F}}\;
\left( \pd{\ov{\cal{H}}}{{\bf E}},\; \pd{\ov{\cal{H}}}{{\bf B}} \right). 
\label{eq:PM_var}
\end{equation}

We begin with an expression for the variation of the reduced Lagrangian density
\begin{eqnarray}
\delta\ov{\cal{L}} & = & \partial_{\mu}\ov{\cal{J}}^{\mu} \;-\; \sum\;\int d^{4}\ov{p}\;\ov{\cal{S}}\;\left\{ \frac{}{} 
\ov{\cal{F}},\; \ov{\cal{H}} \right\}_{\epsilon} \label{eq:delLag_def} \\
 &  &+\; \frac{\delta A_{\nu}}{4\pi} \left[ \pd{}{x^{\mu}} \left( \frac{}{} F^{\mu\nu} - 4\pi\,K^{\mu\nu} \right) 
+ \frac{4\pi}{c}\;\ov{J}^{\nu} \right], \nonumber
\end{eqnarray}
which is generated by the four-potential variation $\delta A_{\nu}$ and the Eulerian variation \cite{Brizard_2000} for the reduced Vlasov distribution 
$\delta\ov{\cal{F}} \equiv -\; \delta\ov{\cal{Z}}^{a}\;\partial_{a}\ov{\cal{F}} \equiv \{ \ov{\cal{S}},\; \ov{\cal{F}} \}_{\epsilon}$, where $\ov{\cal{S}}$ is the generating scalar field for a virtual displacement on reduced phase space, $\delta\ov{\cal{Z}}^{a} \equiv \{ \ov{\cal{Z}}^{a},\;\ov{\cal{S}} \}_{\epsilon}$. Note that the divergence term $\partial_{\mu}\ov{\cal{J}}^{\mu}$, where the reduced Noether four-density is 
\begin{equation}
\ov{\cal{J}}^{\mu} \;\equiv\; \sum\;\int d^{4}\ov{p}\; \ov{\cal{S}}\,\ov{\cal{F}}\;\frac{d_{\epsilon}\ov{x}^{\mu}}{dt} 
\;+\; \frac{\delta A_{\nu}}{4\pi} \left( \frac{}{} F^{\nu\mu} \;-\; 4\pi\, K^{\nu\mu} \right),
\label{eq:Noether_def}
\end{equation}
does not contribute to the variational principle $\int d^{4}x\,\delta\ov{\cal{L}} = 0$ but instead is used to derive exact conservation laws by applications of the Noether method. 

Next, as a result of the variational principle, where the variations $\ov{\cal{S}}$ and $\delta A_{\nu}$ are 
arbitrary (but are required to vanish on the integration boundaries), we obtain the reduced Vlasov equation 
(\ref{eq:Vlasov_epsilon}), $\{ \ov{\cal{F}},\; \ov{\cal{H}}\}_{\epsilon} = 0$, and the reduced ({\it macroscopic}) Maxwell equations
\begin{equation}
\pd{}{x^{\mu}} \left( F^{\mu\nu} \;-\; 4\pi\,K^{\mu\nu} \right) \;=\; -\;\frac{4\pi}{c}\;\ov{J}^{\nu},
\label{eq:Maxwell_var}
\end{equation}
from which we recover the reduced Maxwell equations (\ref{eq:D_Poisson}) and (\ref{eq:H_Ampere}). Here, the polarization-magnetization four-current is expressed in terms of the tensor (\ref{eq:redH_var}) as
\begin{eqnarray*} 
-\;\partial_{\mu}K^{\mu\nu} & = & (-\,\nabla\bdot{\bf P}_{\epsilon}, c^{-1}\partial_{t}{\bf P}_{\epsilon} + 
\nabla\btimes{\bf M}_{\epsilon}) \\
 & \equiv & \left( \rho_{{\rm pol}},\; c^{-1}{\bf J}_{{\rm pol}} + c^{-1}{\bf J}_{{\rm mag}} \right).
\end{eqnarray*}
Note that the electromagnetic field tensor also satisfies $\partial_{\sigma}F_{\mu\nu} + \partial_{\mu}F_{\nu\sigma} + \partial_{\nu}F_{\sigma\mu} = 0$. In addition, we note that the reduced electric-dipole and reduced magnetic-dipole moments (\ref{eq:pi_def}) and (\ref{eq:mu_def}) are also expressed in terms of derivatives of the reduced Hamiltonian $\ov{H}$ as
\[ \vb{\pi}_{\epsilon} \;\equiv\; -\;\pd{\ov{H}}{{\bf E}} \;\;\;{\rm and}\;\;\; \vb{\mu}_{\epsilon} \;\equiv\; -\; \left( \pd{\ov{H}}{{\bf B}} \;+\; \frac{1}{c}\,\frac{d_{\epsilon}\ov{{\bf x}}}{dt}\btimes\pd{\ov{H}}{{\bf E}} \right), \]
which provides a useful consistency check on the reduced Vlasov-Maxwell equations. Lastly, the reduced charge conservation law 
\begin{equation}
0 \;=\; \pd{\ov{J}^{\nu}}{x^{\nu}} \;\equiv\; \pd{\ov{\rho}}{t} \;+\; \nabla\bdot\ov{{\bf J}}
\label{eq:redcon}
\end{equation} 
follows immediately from the reduced Maxwell equations (\ref{eq:Maxwell_var}) as a result of the antisymmetry of 
$F^{\mu\nu}$ and $K^{\mu\nu}$ (i.e., $\partial_{\mu\nu}^{2}F^{\mu\nu} \equiv 0 \equiv \partial_{\mu\nu}^{2}K^{\mu\nu}$).

\subsection{Reduced Energy-momentum Conservation Laws}

We now derive the energy-momentum conservation law from the reduced Noether equation $\delta\ov{\cal{L}} \equiv 
\partial_{\mu}\ov{\cal{J}}^{\mu}$ associated with space-time translations generated by $\delta x^{\sigma} \equiv (c\,\delta t, \delta{\bf x})$, where the variations
$(\ov{\cal{S}}, \delta A_{\nu}, \delta\ov{\cal{L}})$ are
\begin{equation}
\left. \begin{array}{rcl}
\ov{\cal{S}} & = & \ov{p}_{\sigma}\,\delta x^{\sigma} \\
 &  & \\
\delta A_{\nu} & = & F_{\nu\sigma}\,\delta x^{\sigma} \;-\; \partial_{\nu}\left( A_{\sigma}\,\delta x^{\sigma} \right) \\
 &  & \\
\delta\ov{\cal{L}} & = & -\;\partial_{\sigma} \left( \ov{\cal{L}}\;\delta x^{\sigma} \right)
\end{array} \right\}.
\label{eq:noether_var}
\end{equation}
Here, we note that $\delta x^{\nu} \equiv \{ \ov{x}^{\nu},\; \ov{\cal{S}}\}_{\epsilon}$, while the variations 
$\delta A_{\nu}$ and $\delta\ov{\cal{L}}$ are expressed in terms of the Lie derivative $\pounds_{\delta x}$ generated by 
$\delta x^{\nu}$ as $\delta A_{\nu}\,\exd x^{\nu} \equiv -\;\pounds_{\delta x}(A_{\nu}\,\exd x^{\nu})$ and $\delta\ov{\cal{L}}\,\exd^{8}\ov{\cal{Z}} \equiv -\;\pounds_{\delta x}(\ov{\cal{L}}\,\exd^{8}\ov{\cal{Z}})$.

After some cancellations introduced through the use of the reduced Maxwell equations (\ref{eq:Maxwell_var}), we obtain the reduced energy-momentum conservation law $\partial_{\mu}T^{\mu\nu} \equiv 0$, where the reduced energy-momentum tensor is defined as
\begin{eqnarray}
T^{\mu\nu} & \equiv & \frac{g^{\mu\nu}}{16\pi}\; {\sf F}:{\sf F} \;-\; \frac{1}{4\pi} \left( F^{\mu\sigma} \;-\; 4\pi\,
K^{\mu\sigma} \right) F_{\sigma}^{\;\;\nu} \nonumber \\
 &  &+\; \sum \int d^{4}\ov{p}\;\ov{\cal{F}} \left( \frac{d_{\epsilon}\ov{x}^{\mu}}{dt}\;\ov{p}^{\nu} \right) \;-\; 
\frac{1}{c}\,\ov{J}^{\mu}A^{\nu}.
\label{eq:Tmunu_def}
\end{eqnarray}
Here, the antisymmetric tensor $K^{\mu\sigma}$, defined in Eq.~(\ref{eq:redH_var}), represents the effects of reduced polarization and magnetization. Note that, while the last two terms are individually gauge-dependent, their sum is invariant under the gauge transformation (\ref{eq:gauge}). Explicit proofs of energy-momentum conservation for the reduced Vlasov-Maxwell equations based on the reduced energy-momentum tensor (\ref{eq:Tmunu_def}) are presented in Appendix \ref{sec:App_A}. Lastly, additional angular-momentum conservation laws can be derived from the reduced Noether equation $\delta\ov{\cal{L}} = \partial_{\mu}\ov{\cal{J}}^{\mu}$ by considering invariance of the reduced Lagrangian density $\ov{\cal{L}}$ with respect to arbitrary rotations in space.

\section{\label{sec:Sum}Summary}

In this paper, the general theory for the reduced Vlasov-Maxwell equations was presented based on the asymptotic elimination of fast time scales by Lie-transform Hamiltonian perturbation method. This dynamical reduction is based on a near-identity transformation on extended phase space, which induces transformations on the Vlasov distribution and the Vlasov operator, as well as introducing a natural (push-forward) representation of charge-current densities in terms of reduced charge-current densities and their associated reduced polarization and magnetization effects. The variational formulation of the reduced Vlasov-Maxwell equations allows the derivation of exact energy-momentum conservation laws by Noether method.

The Table shown below summarizes the polarization and magnetization effects observed in reduced Vlasov-Maxwell equations that have important applications in plasma physics.

\begin{widetext}

\[ \begin{array}{|c|c|c|} \hline
\;\;\mbox{Reduced Dynamics}\;\; & \;\;\;\;\vb{\pi}_{\epsilon}\;\;\;\; & \;\;\;\;\vb{\mu}_{\epsilon}\;\;\;\; \\ \hline \hline
 &  & \\
\mbox{Guiding-center}  & (mc^{2}/B^{2})\;{\bf E}_{\bot} \;+\; e\,\bhat\btimes
{\bf v}_{{\rm B}}/\Omega & -\,\ov{\mu}\;\bhat  \\ 
 &  & \\ \hline
 &  & \\
\mbox{Gyrocenter}  & \epsilon\;(c\bhat_{0}/B_{0})\btimes\left( e\,{\bf A}_{1}/c \;+\; m\,{\bf u}_{{\rm E}1} \;+\; \ov{p}_{\|}\,{\bf B}_{1}/B_{0}\right) & -\,\ov{\mu}\;(\bhat_{0} + \epsilon\,{\bf B}_{1}/B_{0})  \\ 
 &  & \\ \hline
 &  & \\
\mbox{Oscillation-center} & \epsilon^{2}\,e{\bf k}\btimes(-i\,\wt{\vb{\xi}}^{*}\btimes\wt{\vb{\xi}}) & \epsilon^{2}\,e\omega^{\prime}/c\;(-i\,
\wt{\vb{\xi}}^{*}\btimes\wt{\vb{\xi}}) \\ 
 &  & \\ \hline
\end{array} \]

\end{widetext}

First, in guiding-center Hamiltonian theory \cite{RGL_83,ANK_86,Brizard_95} for a strongly magnetized plasma in the presence of a background electric field ${\bf E} \equiv -\,\nabla\Phi$, the fast time scale is associated with the rapid gyromotion of a charged particle about a magnetic field line and fast-time-averaging is carried out by averaging with respect to the gyroangle. The fast-time-averaged reduced electric-dipole moment $\ov{\vb{\pi}}_{{\rm gc}} \equiv e\,
\ov{\vb{\rho}}_{{\rm gc}}$ includes effects due to the background electric field $({\bf v}_{{\rm E}} = c\,{\bf E}\btimes
{\bf B}/B^{2})$ as well as the magnetic ($\nabla$B and curvature) drift velocity ${\bf v}_{{\rm B}}$. The fast-time-averaged particle polarization velocity $d_{{\rm gc}}\ov{\vb{\rho}}_{{\rm gc}}/dt$ includes the standard polarization drift velocity $(c/B\Omega)\,\partial{\bf E}_{\bot}/\partial t$, which is not included in the guiding-center drift velocity $d_{{\rm gc}}\ov{{\bf x}}/dt \equiv {\bf v}_{{\rm gc}} = {\bf v}_{{\rm E}} + {\bf v}_{{\rm B}}$. On the other hand, the reduced magnetic-dipole moment yields the classical parallel magnetization term $\ov{\vb{\mu}}_{{\rm gc}} \equiv -\,\ov{\mu}\,\bhat$ (where $\ov{\mu}$ denotes the guiding-center magnetic-moment adiabatic invariant and $\bhat \equiv {\bf B}/B$ denotes the unit vector along a magnetic-field line), which enables the reconciliation of the particle current ${\bf J}$ with the guiding-center current ${\bf J}_{{\rm gc}}$ through the relation ${\bf J} \equiv {\bf J}_{{\rm gc}} + 
{\bf J}_{{\rm mag}}$ (valid in a static magnetized plasma).

Next, gyrocenter Hamiltonian theory \cite{BH_06} describes the reduced (gyroangle-independent) perturbed guiding-center Hamiltonian dynamics associated with low-frequency, electric and magnetic fluctuations $(\epsilon\,{\bf E}_{1}, \epsilon\,{\bf B}_{1})$ in a strongly-magnetized plasma (with static magnetic field 
${\bf B}_{0} \equiv B_{0}\,\bhat_{0}$ and ${\bf E}_{0} \equiv 0$). Note that the results shown here are valid only in the limit of zero Larmor radius \cite{BH_06}. The reduced electric-dipole moment includes not only the perturbed polarization-drift term ($mc\,\bhat_{0}/B_{0}\btimes{\bf u}_{{\rm E}1}$), but also the effects due to magnetic flutter $(c\ov{p}_{\|}\,\bhat_{0}\btimes{\bf B}_{1}/B_{0})$ and the inductive part of the perturbed $E\times B$ velocity (i.e., the polarization drift velocity includes the higher-order correction $-\partial_{t}{\bf A}_{1}\btimes\bhat_{0}/B_{0}$ to the perturbed $E\times B$ velocity $-\nabla\Phi_{1}\btimes c\,\bhat_{0}/B_{0}$). On the other hand, the reduced magnetic-dipole moment includes a correction to the classical parallel magnetization term due to the perturbed magnetic field: 
$\ov{\vb{\mu}}_{{\rm gy}} \equiv -\,\ov{\mu}\,(\bhat_{0} + \epsilon\,{\bf B}_{1}/B_{0})$, where $\ov{\mu}$ now denotes the gyrocenter magnetic-moment adiabatic invariant.

Lastly, oscillation-center Hamiltonian theory \cite{CK_81} describes the reduced dynamics of charged particles interacting with a high-frequency electromagnetic wave in a weakly-inhomogeneous plasma for which the eikonal approximation is valid. The eikonal representation for the wave fields is $({\bf E}_{1}, {\bf B}_{1}) \equiv 
(\wt{{\bf E}}_{1}, \wt{{\bf B}}_{1})\;\exp(i\epsilon_{0}^{-1}\Theta) + {\rm c.c.}$, where $\epsilon_{0} \ll 1$ denotes the eikonal small parameter while the eikonal phase $\Theta(\epsilon_{0}{\bf r}, \epsilon_{0}t)$ is used to define $\omega \equiv -\epsilon_{0}^{-1}\partial_{t}\Theta$ and ${\bf k} \equiv \epsilon_{0}^{-1}\nabla\Theta$, with $\omega^{\prime} \equiv \omega - {\bf k}\bdot{\bf v}$ denoting the Doppler-shifted wave frequency. Note that fast-time-averaging, here, is carried out by averaging with respect to the eikonal phase $\Theta$. By considering the simplest case of an unmagnetized plasma \cite{CK_81}, the first-order term for the displacement $\vb{\rho}_{\epsilon} = \epsilon\,\vb{\xi} + \cdots$ has the eikonal amplitude
\[ \wt{\vb{\xi}} \;=\; -\;\frac{e}{m\omega^{\prime 2}} \left( \wt{{\bf E}}_{1} \;+\; \frac{{\bf v}}{c}\btimes\wt{{\bf B}}_{1} \right), \]
where $\epsilon \ll 1$ denotes the amplitude of the electromagnetic wave represented by the first-order fields ${\bf E}_{1}$ and ${\bf B}_{1}$.  Note that both reduced oscillation-center electric-dipole and magnetic-dipole moments are quadratic functions of the wave fields. An additional wave-action conservation law results from the invariance of the reduced Lagrangian density on the eikonal phase but its derivation is outside the scope of this work.

\acknowledgments

The present work was supported by the National Science Foundation under grant number DMS-0317339.

\appendix

\section{\label{sec:App_A}Explicit Proofs of Energy-momentum Conservation for the Reduced Vlasov-Maxwell Equations}

In this Appendix, we present explicit proofs of energy-momentum conservation based on the reduced energy-momentum stress tensor (\ref{eq:Tmunu_def}). We begin with the reduced energy conservation law $\partial\cal{E}/\partial t + \nabla\bdot{\bf S} = 0$, where the reduced energy density $\cal{E} \equiv T^{00}$ is
\begin{eqnarray*}
\cal{E} & = & -\;\frac{1}{8\pi}\,\left( |{\bf E}|^{2} \;-\; |{\bf B}|^{2} \right) \;+\; \frac{{\bf D}\bdot{\bf E}}{4\pi} \\
 &  &+\; \sum \int d^{3}\ov{p}\;\ov{f}\;\ov{H} \;-\; \ov{\rho}\,\Phi,
\end{eqnarray*}
where the $\ov{w}$-integration was performed. First, using the reduced charge conservation law (\ref{eq:redcon}) and the identity
\begin{eqnarray*} 
\sum \int d^{3}\ov{p}\; \ov{f}\;\pd{\ov{H}}{t} & \equiv & \ov{\rho}\;\pd{\Phi}{t} \;-\; \frac{\ov{{\bf J}}}{c}\bdot
\pd{{\bf A}}{t} \\
 &  &-\; {\bf P}_{\epsilon}\bdot\pd{{\bf E}}{t} \;-\; {\bf M}_{\epsilon}\bdot\pd{{\bf B}}{t},
\end{eqnarray*}
we obtain
\begin{eqnarray*}
\pd{\cal{E}}{t} & = & \frac{{\bf E}}{4\pi}\bdot\pd{{\bf D}}{t} \;+\; \frac{{\bf H}}{4\pi}\bdot\pd{{\bf B}}{t} \;+\; 
{\bf E}\bdot\ov{{\bf J}} \\
 &  &+\; \nabla\bdot\left( \ov{{\bf J}}\;\Phi \right) \;-\; \sum \int d^{3}\ov{p}\; \ov{H}\;\{ \ov{f},\; \ov{H}\},
\end{eqnarray*}
where we used the reduced Vlasov equation (\ref{eq:redVlasov_def}) to obtain the last term. Lastly, using the reduced Maxwell equations (\ref{eq:EB_constraint}) and (\ref{eq:H_Ampere}) and the identity
\begin{eqnarray*} 
\int d^{3}\ov{p}\; \ov{H}\;\{ \ov{f},\; \ov{H}\} & = & \int d^{3}\ov{p}\; \{ \ov{f}\,\ov{H},\; \ov{H}\} \\
 & \equiv & \nabla\bdot\left( \int d^{3}\ov{p}\; \ov{f}\,\ov{H}\;\frac{d_{\epsilon}\ov{{\bf x}}}{dt} \right),
\end{eqnarray*}
we finally obtain
\begin{eqnarray*}
\pd{\cal{E}}{t} & = & -\;\nabla\bdot \left( \frac{c}{4\pi}\,{\bf E}\btimes{\bf H} \;-\; \ov{{\bf J}}\,\Phi \right. \\
 &  &\left. +\; \sum \int d^{3}\ov{p}\;\ov{f}\,\ov{H}\;\frac{d_{\epsilon}\ov{{\bf x}}}{dt} \right) \;\equiv\; -\;\nabla\bdot{\bf S},
\end{eqnarray*}
where the energy-density flux $S^{i} \equiv c\,T^{i0}$. 

Next, we consider the reduced momentum conservation law $\partial\vb{\Pi}/\partial t + \nabla\bdot{\sf T} = 0$, where the reduced momentum density $\Pi^{i} \equiv T^{0i}/c$ is
\[ \vb{\Pi} \;=\; \frac{{\bf D}\btimes{\bf B}}{4\pi\,c} \;-\; \frac{\ov{\rho}}{c}\;{\bf A} \;+\; \sum \int d^{3}\ov{p}\;
\ov{f}\;\ov{{\bf p}}, \]
which has the Minkowski form \cite{Jackson}. By substituting the reduced Vlasov-Maxwell equations (\ref{eq:EB_constraint}) and (\ref{eq:D_Poisson})-(\ref{eq:H_Ampere}) and the reduced charge conservation law (\ref{eq:redcon}), we obtain
\begin{eqnarray*}
\pd{\vb{\Pi}}{t} & = & \nabla\bdot\left[\; \frac{1}{4\pi} \left( {\bf B}\,{\bf H} \;+\; {\bf D}\,{\bf E} \right) \;+\; \frac{1}{c}\;\ov{{\bf J}}\,{\bf A} \right. \\
 &  &\left.-\; \sum \int d^{3}\ov{p}\;\ov{f}\;\left( \frac{d_{\epsilon}\ov{{\bf x}}}{dt}\;\ov{{\bf p}} \right) \;\right] \\
 &  &\mbox{}-\; \sum \int d^{3}\ov{p}\;\ov{f}\;\nabla\ov{H} \;+\; \nabla\Phi\;\ov{\rho} \;-\; \nabla{\bf A}\bdot
\frac{\ov{{\bf J}}}{c} \\
 &  &-\; \left( \nabla{\bf H}\bdot{\bf B} \;+\; \nabla{\bf E}\bdot{\bf D} \right).
\end{eqnarray*}
Lastly, using the identities
\begin{eqnarray*} 
\sum \int d^{3}\ov{p}\; \ov{f}\;\nabla\ov{H} & \equiv & \nabla\Phi\;\ov{\rho} \;-\; \nabla{\bf A}\bdot
\frac{\ov{{\bf J}}}{c} \\
 &  &-\; \nabla{\bf E}\bdot{\bf P}_{\epsilon} \;-\; \nabla{\bf B}\bdot{\bf M}_{\epsilon},
\end{eqnarray*}
and
\begin{eqnarray*} 
\nabla{\bf H}\bdot{\bf B} \;+\; \nabla{\bf E}\bdot{\bf D} & = & \nabla\left( \frac{|{\bf E}|^{2}}{2} \;+\; 
\frac{|{\bf B}|^{2}}{2} \;-\; 4\pi\,{\bf M}_{\epsilon}\bdot{\bf B} \right) \\
 &  &+\; 4\pi \left(\nabla{\bf E}\bdot{\bf P}_{\epsilon} \;+\; \nabla{\bf B}\bdot{\bf M}_{\epsilon} \right),
\end{eqnarray*}
we finaly obtain
\begin{eqnarray*}
\pd{\vb{\Pi}}{t} & = & \nabla\bdot\left[ \frac{1}{4\pi} \left( {\bf D}\,{\bf E} + {\bf B}\,{\bf H} \right) -
\sum \int d^{3}\ov{p}\,\ov{f}\left( \frac{d_{\epsilon}\ov{{\bf x}}}{dt}\ov{{\bf p}} \right) \right. \\
 &  &\left.+ \frac{1}{c}\;\ov{{\bf J}}\,{\bf A} - \frac{{\bf I}}{4\pi} \left( \frac{|{\bf E}|^{2}}{2} + 
\frac{|{\bf B}|^{2}}{2} - 4\pi\,{\bf M}_{\epsilon}\bdot{\bf B} \right) \right] \\
 & \equiv & -\;\nabla\bdot{\sf T}.
\end{eqnarray*}

\end{document}